# Global Impact of COVID-19 Restrictions on the Atmospheric Concentrations of Nitrogen Dioxide and Ozone

Christoph A. Keller[1,2], Mat. J. Evans[3,4], K. Emma Knowland[1,2], Christa A. Hasenkopf[5], Sruti Modekurty[5], Robert A. Lucchesi[1,6], Tomohiro Oda[1,2], Bruno B. Franca[7], Felipe C. Mandarino[7], M. Valeria Díaz Suárez[8], Robert G. Ryan[9], Luke H. Fakes[3,4], Steven Pawson[1]

[1]NASA Global Modeling and Assimilation Office, Goddard Space Flight Center, Greenbelt, MD, USA
[2]Universities Space Research Association, Columbia, MD, USA
[3]Wolfson Atmospheric Chemistry Laboratories, Department of Chemistry, University of York, York, YO10 5DD, UK
[4]National Centre for Atmospheric Science, University of York, York, YO10 5DD, UK
[5]OpenAQ, Washington, DC, USA
[6]Science Systems and Applications, Inc., Lanham, MD, USA
[7]Municipal Government of Rio de Janeiro, Rio de Janeiro, Brazil
[8]Secretaria de Ambiente, Quito, Ecuador
[9]School of Earth Sciences, The University of Melbourne, Australia

*Correspondence to*: Christoph A. Keller (christoph.a.keller@nasa.gov)

**Abstract.** Social-distancing to combat the COVID-19 pandemic has led to widespread reductions in air pollutant emissions. Quantifying these changes requires a business as usual counterfactual that accounts for the synoptic and seasonal variability of air pollutants. We use a machine learning algorithm driven by information from the NASA GEOS-CF model to assess changes in nitrogen dioxide ($NO_2$) and ozone ($O_3$) at 5,756 observation sites in 46 countries from January through June 2020. Reductions in $NO_2$ correlate with timing and intensity of COVID-19 restrictions, ranging from 60% in severely affected cities (e.g., Wuhan, Milan) to little change (e.g., Rio de Janeiro, Taipei). On average, $NO_2$ concentrations were 18% lower than business as usual from February 2020 onward. China experienced the earliest and steepest decline, but concentrations since April have mostly recovered and remained within 5% to the business as usual estimate. $NO_2$ reductions in Europe and the US have been more gradual with a halting recovery starting in late March. We estimate that the global $NO_x$ ($NO+NO_2$) emission reduction during the first 6 months of 2020 amounted to 2.9 TgN, equivalent to 5.1% of the annual anthropogenic total. The response of surface $O_3$ is complicated by competing influences of non-linear atmospheric chemistry. While surface $O_3$ increased by up to 50% in some locations, we find the overall net impact on daily average $O_3$ between February - June 2020 to be small. However, our analysis indicates a flattening of the $O_3$ diurnal cycle with an increase in night time ozone due to reduced titration and a decrease in daytime ozone, reflecting a reduction in photochemical production.
The $O_3$ response is dependent on season, time scale, and environment, with declines in surface $O_3$ forecasted if $NO_x$ emission reductions continue.

# 1 Introduction

The stay-at-home orders imposed in many countries during the Northern Hemisphere spring of 2020 to slow the spread of the severe acute respiratory syndrome coronavirus 2 (SARS-CoV-2, hereafter COVID-19), led to a sharp decline in human activities across the globe (Le Quéré et al., 2020). The associated decrease in industrial production, energy consumption, and transportation resulted in a reduction in the emissions of air pollutants, notably nitrogen dioxide ($NO_2$) (Liu et al., 2020a; Dantas et al., 2020; Petetin et al., 2020; Tobias et al., 2020; Le et al., 2020). Nitrogen oxides ($NO_x$=NO+$NO_2$) have a short atmospheric lifetime and are predominantly emitted during the combustion of fossil fuel for industry, transport and domestic activities (Streets et al., 2013, Duncan et al., 2016). Atmospheric $NO_2$ concentrations thus readily respond to local changes in $NO_x$ emissions (Lamsal et al., 2011). While this may provide both air quality and climate benefits, a quantitative assessment of the magnitude of these impacts is complicated by the natural variability of air pollution due to variations in synoptic conditions (weather), seasonal effects, and long-term emission trends as well as the non-linear responses between emissions and concentrations. Thus, simply comparing the concentration of pollutants during the COVID-19 period to those immediately before or to the same period in previous years is not sufficient to indicate causality. An emerging approach to address this problem is to develop machine-learning based 'weather-normalization' algorithms to establish the relationship between local meteorology and air pollutant surface concentrations (Grange et al., 2018; Grange and Carslaw, 2019; Petetin et al., 2020). By removing the meteorological influence, these studies have tried to better quantify emission changes as a result of a perturbation.

Here we adapt this weather-normalization approach to not only include meteorological information but also compositional information in the form of the concentrations and emissions of chemical constituents. Using a collection of surface observations of $NO_2$ and ozone ($O_3$) from across the world from 2018 to present (Section 2.1), we develop a 'bias-correction' methodology for the NASA global atmospheric composition model GEOS-CF (Section 2.2) which corrects the model output at each observational site based on the observations for 2018 and 2019 (Section 2.3). These biases reflect errors in emission estimates, sub-gridscale local influences (representational error), or meteorology and chemistry. Since the GEOS-CF model makes no adjustments to the anthropogenic emissions in 2020, and no 2020 observations are included in the training of the bias corrector, the bias-corrected model (hereafter BCM) predictions for 2020 represents a business as usual scenario at each observation site that can be compared against the actual observations. This allows the impact of COVID-19 containment measures on air quality to be explored, taking into account meteorology and the long-range transport of pollutants. We first apply this to the concentration of $NO_2$ (Section 3.1), and then $O_3$ (Section 3.2) and explore the differences between the counterfactual prediction and the observed concentrations. In Section 3.3 we explore how the observed changes in the $NO_2$ concentrations relate to emission of $NO_x$. Finally, in Section 4 we discuss our conclusions and speculate what the COVID-19 restrictions might mean for the second half of 2020.

## 2 Methods

### 2.1 Observations

Our analysis builds on the recent development of unprecedented public access to air pollution model output and air quality observations in near real-time. We compile an air quality dataset of hourly surface observations for a total of 5,756 sites (4,778 for $NO_2$ and 4,463 for $O_3$) in 46 countries for the time period January 1, 2018 to July 1, 2020, as summarized in Fig. 1 and Table 1. The vast majority of the observations were obtained from the OpenAQ platform and the air quality data portal of the European Environment Agency (EEA). Both platforms provide harmonized air quality observations in near real-time, greatly facilitating the analysis of otherwise disparate data sources. For Japan, we obtained hourly surface observations for a total of 225 sites in Hokkaido, Osaka, and Tokyo from the Atmospheric Environmental Regional Observation System (AEROS) (MOE, 2020). To improve data coverage in under-sampled regions, we further included observations from the cities of Rio de Janeiro (Brazil), Quito (Ecuador), and Melbourne (Australia). All cities offer continuous, hourly observations of $NO_2$ and $O_3$ over the full analysis period, thus offering an excellent snapshot of air quality at these locations. We include all sites with at least 365 days of observations between Jan 1, 2018 and December 31, 2019, and an overall data coverage of 75% or more since the first day of availability. The final $NO_2$ and $O_3$ dataset comprise $8.9 \times 10^7$ and $8.2 \times 10^7$ hourly observations, respectively.

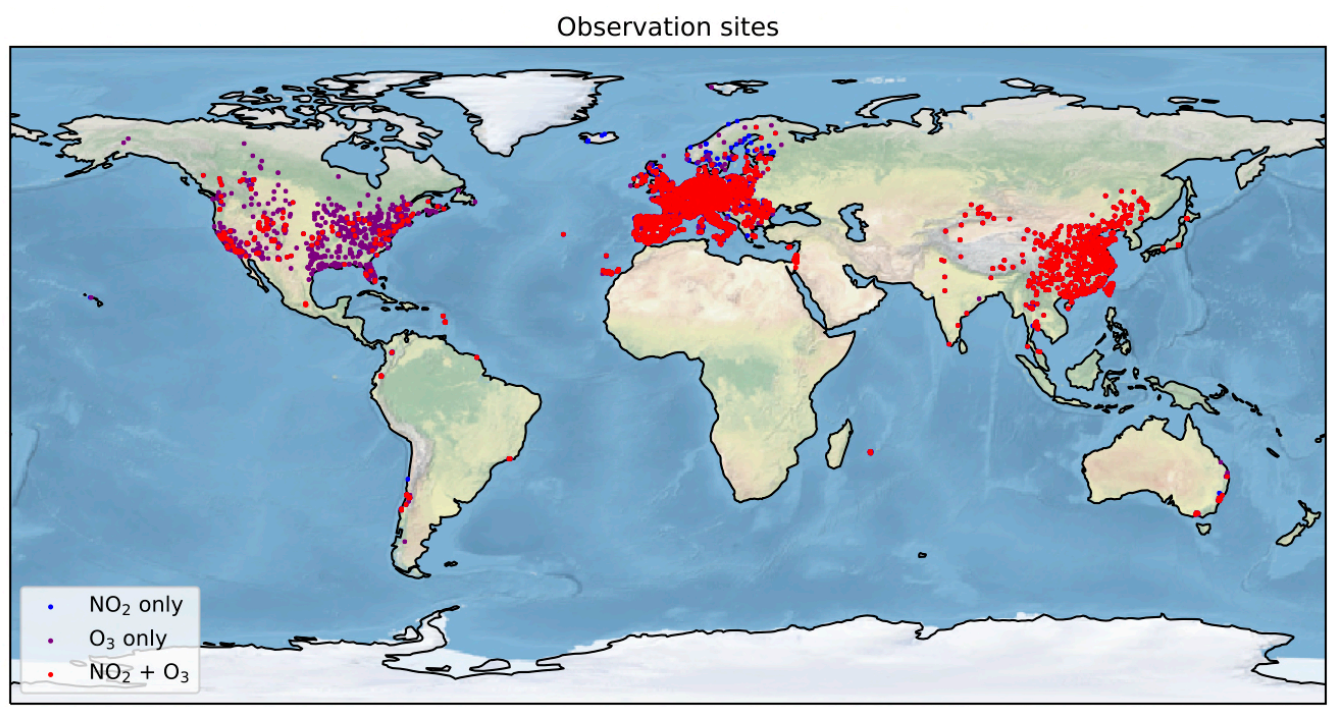

**Figure 1:** Location of the 5,756 observation sites included in the analysis. Red points indicate sites with both $NO_2$ and $O_3$ observations (3,485 in total), purple points show locations with $O_3$ observations only (978 sites) and blue points show locations with $NO_2$ observations only (1,293 sites).

Table 1: Observational data sources used in the analysis. Time period covers Jan 1, 2018 - July 1, 2020.

| Name | Countries | Sites | Source |
|---|---|---|---|
| OpenAQ | Australia, China, India, Hong Kong, Taiwan, Thailand, Canada, Chile, Colombia, United States | 2410 | https://openaq.org/ |
| EEA | Austria, Belgium, Bosnia and Herzegovina, Bulgaria, Croatia, Cyprus, Czech Republic, Denmark, Estonia, Finland, France, Germany, Greece, Hungary, Iceland, Ireland, Italy, Latvia, Lithuania, Luxembourg, Macedonia, Malta, Netherlands, Norway, Poland, Portugal, Romania, Serbia, Slovakia, Slovenia, Spain, Sweden, Switzerland, United Kingdom, | 3101 | https://discomap.eea.europa.eu/map/fme/AirQualityExport.htm |
| AEROS | Japan | 225 | http://soramame.taiki.go.jp/Index.php |
| EPA Victoria | Australia (Melbourne) | 4 | http://sciwebsvc.epa.vic.gov.au/aqapi/Help |
| Secretaria de Ambiente, Quito | Ecuador (Quito) | 8 | http://www.quitoambiente.gob.ec/ambiente/index.php/datos-horarios-historicos |
| Municipal Government of Rio de Janeiro | Brazil (Rio de Janeiro) | 8 | http://www.data.rio/datasets/dados-hor%C3%A1rios-do-monitoramento-da-qualidade-do-ar-monitorar |

**2.2 Model**

Meteorological and atmospheric chemistry information at each of the air quality observation sites is obtained from the NASA Goddard Earth Observing System Composition Forecast (GEOS-CF) model. GEOS-CF integrates the GEOS-Chem atmospheric chemistry model (v12-01) into the GEOS Earth System Model (Long et al., 2015; Hu et al., 2018) and provides global hourly analyses of atmospheric composition at 25x25 km$^2$ spatial resolution, available in near real-time at https://gmao.gsfc.nasa.gov/weather_prediction/GEOS-CF/data_access/ (Knowland et al., 2020). Anthropogenic emissions are prescribed using monthly Hemispheric Transport of Air Pollution (HTAP) bottom-up emissions (Janssens-Maenhout et al., 2015), scaled from 2010 to 2018 based on annual average NO$_2$ satellite observations from the NASA Aura Ozone Monitoring Instrument (OMI) (Boersma et al., 2011). The 2018 anthropogenic emissions are subsequently used for years 2019 and 2020. Therefore, GEOS-CF does not account for any anthropogenic emission changes since 2018, notably any anthropogenic emission reductions related to COVID-19 restrictions. However, it does capture the variability in natural emissions such as wildfires (based on the Quick Fire Emissions Dataset, QFED) (Darmenov and Da Silva, 2015), lightning and biogenic emissions (Keller et al., 2014). While the meteorology and stratospheric ozone in GEOS-CF are fully constrained by pre-computed analysis fields produced by other GEOS systems (Lucchesi, 2015; Wargan et al., 2015), no trace-gas observations are directly assimilated into the current version of GEOS-CF. It thus provides a "business as usual" estimate of NO$_2$ and O$_3$ that can be used as a baseline for input into the meteorological normalization process.

## 2.3 Machine learning bias correction

We use the XGBoost machine learning algorithm (https://xgboost.readthedocs.io/en/latest/#) (Chen and Guestrin, 2016; Frery et al., 2017) to develop a machine learning model to predict the time-varying bias at each observation site. XGBoost uses the Gradient Boosting framework to build an ensemble of decision trees, trained iteratively on the residual errors to stage-wise improve the model predictions (Friedman, 2001). Based on the 2018-2019 observation-model differences, the machine learning model is trained to predict the systematic (recurring) model bias between hourly observations and the co-located model predictions. These biases can be due to errors in the model, such as emission estimates, sub-gridscale local influences (representational error), or meteorology and chemistry. At each site, we use 50% of the observations to train the model (randomly selected), and the other half is used for validation. Since the machine learning algorithm can be sensitive to outliers, all observations below or above 2 standard deviations from the mean were removed from the analysis. Figures 2 and 3 summarize the machine learning model statistics for $NO_2$ and $O_3$, respectively.

The input variables fed into the XGBoost algorithm encompass 8 meteorological parameters (as simulated by the GEOS-CF model: surface north- and eastward wind components, surface temperature, surface relative humidity, total cloud coverage, total precipitation, surface pressure, and planetary boundary layer height), modelled surface concentrations of 49 chemical species ($O_3$, $NO_x$, carbon monoxide, VOCs, and aerosols), and 31 modelled emissions at the given location. In addition, we provide as input features the hour-of-day, day of week, and month of the year; these allow the machine learning model to identify systematic observation-model mismatches related to the diurnal, weekly and seasonal cycle of the pollutants. In addition, for sites with observations available for the full two years, we provide the calendar days since Jan 1, 2018 as an additional input feature to also correct for inter-annual trends in air pollution, e.g., due to a steady decrease in emissions not captured by the model. This follows a similar technique to Ivatt and Evans (2020) and Petetin et al. (2020).

Based on the machine-learning predictor trained on 2018-2019, we predict the model bias for the observation sites in 2020 and adjust the GEOS-CF model predicted concentrations accordingly.

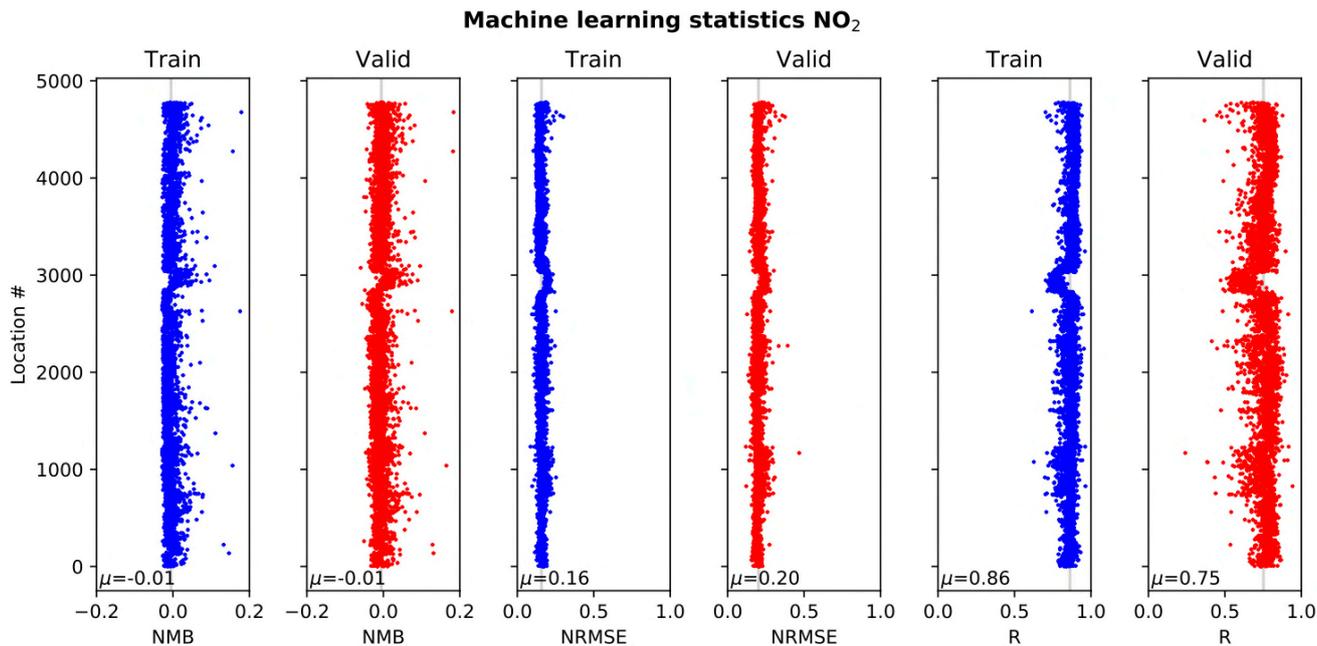

**Figure 2:** Machine learning statistics between hourly observations and the corresponding bias-corrected model predictions for each observation location. Shown are the normalized mean bias (NMB), normalized root mean square error (NRMSE) and Pearson correlation coefficient (R) for the training data (blue) and the validation data (red). Grey line indicates mean value across all locations. NMB is defined as mean bias normalized by average concentration at the given site, and NRMSE is the root mean square error normalized by the range of the 95-percentile concentration and 5-percentile concentration.

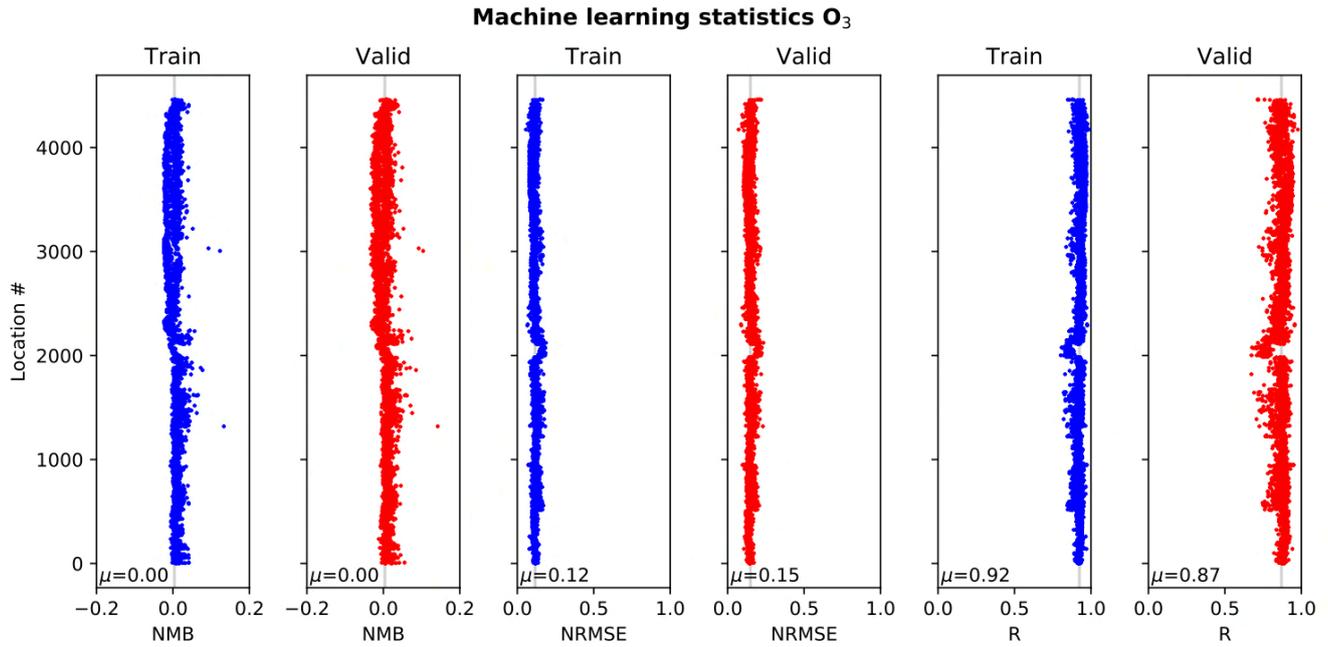

**Figure 3:** As Figure 2 but for $O_3$.

## 3 Results

**3.1 Nitrogen dioxide**

Figure 4 shows the weekly mean observations of $NO_2$ concentration, the GEOS-CF estimate and the BCM prediction for five cities, Wuhan (China), Taipei (Taiwan), Milan (Italy), New York (USA) and Rio de Janeiro (Brazil), from January 2018 through June 2020. While the uncorrected GEOS-CF model estimates fail to represent some aspects of the observations due to uncertainties in model processes and the observation-model scale mismatch, the BCM closely follows the observations for years 2018 and 2019. Even with only half of the 2018 and 2019 observations fed into the machine learning algorithm, the BCM equally captures the observations included in the training and left out for validation (see also Fig. 2).

The grey region in Fig. 4 shows the start and end of the implementation of COVID-19 containment measures. The start and end dates for these are from https://en.wikipedia.org/wiki/COVID-19_pandemic_lockdowns or based on local knowledge. Once containment is implemented, observed concentrations start to diverge from the BCM prediction for Wuhan, Milan and New York (Fig. 4). For Wuhan, we find a reduction in $NO_2$ of 60% relative to the expected BCM value for February and March 2020, and similar decreases are found over Milan (60%) and New York (45%) starting in mid-March and lasting through April (Fig. 4; Tables A1-A3). For cities where restrictions have been mainly removed (Wuhan, Milan) concentrations rise back towards the BCM value, although in neither city are the concentrations fully restored to what might be expected based on the business as usual GEOS-CF simulation.

Looking more broadly at cities around the globe, 50 of the 61 specifically analysed cities feature $NO_2$ reductions of between 20-50% (Fig. A1-A3 and Tables A1-A3). Most locations issued social distancing recommendations prior to the legal lockdowns, and the $NO_2$ declines often precede the official lockdown date by 7-14 days (e.g., Brussels, London, Boston, Phoenix, and Washington, DC).

For Taipei and Rio de Janeiro, the observations and the BCM show little difference (Fig. 4), consistent with the less stringent quarantine measures in these places. Other cities with only short-term $NO_2$ reductions of less than 25% include Atlanta (USA), Budapest (Hungary), and Melbourne (Australia), again correlating with the comparatively relaxed containment measures in these places (Fig. A1-A3). In contrast, Tokyo (Japan) and Stockholm (Sweden), which also implemented a less aggressive COVID-19 response, exhibit $NO_2$ reductions comparable to those of cities with official lockdowns (>20%), suggesting that economic and human activities were similarly subdued in those cities.

Substantial differences exist between cities in South America, with Rio de Janeiro and Santiago de Chile showing little change thus far in 2020, whereas Quito (Ecuador) and Medellin (Colombia) experienced a greater than 50% reduction in $NO_2$ after the initiation of strict restrictions measures in mid-March (Fig. A3 and Table A3). Concentrations in Medellin rebounded sharply in April and May, while concentrations in Quito remained 50% below business as usual throughout most of May and only started to return back to normal values in June.

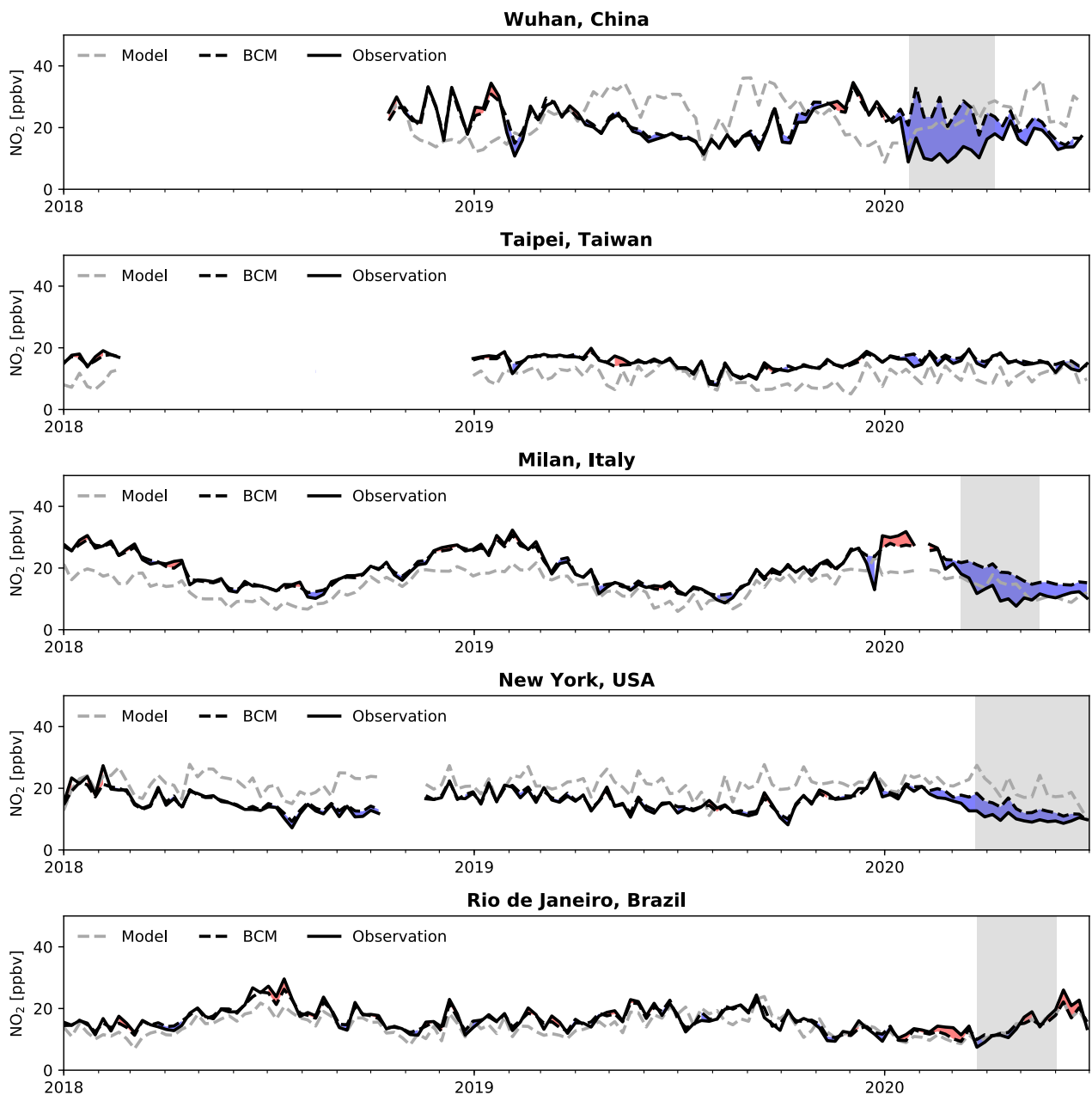

**Figure 4.** Comparison of $NO_2$ surface concentrations (ppbv = nmol mol-1) for Wuhan, Taipei, Milan, New York, and Rio de Janeiro for January 2018 through June 2020. Observed values are shown in solid black, the original GEOS-CF model simulation is shown in dashed grey, and the BCM predictions are in dashed black. The area between observations and BCM predictions is shaded blue (red) if observations are lower (higher) than BCM predictions. Grey areas represent the period of lockdown. Shown are the 7-day average mean values for the 9, 18, 19, 14 and 2 observational sites in Wuhan, Taipei, Milan, New York, and Rio de Janeiro, respectively. Observations for China are only available starting in mid-September 2018.

To evaluate the large-scale impact of COVID-19 restrictions on air quality, we aggregate the individual observation-model comparisons by country. We note that our estimates for some countries (e.g., Brazil, Colombia) are based on a single city and likely not representative of the whole country. On a country level, we find the sharpest and earliest drop in $NO_2$ over China, where observed concentrations fell, on average, 55% below their expected value in early February when restrictions were implemented (Fig. 5). Concentrations remained at this level until late February, at which point they started to increase until restrictions were significantly relaxed in early April. Our analysis suggests that Chinese $NO_2$ concentrations have recovered to within 5% of the business as usual since then. For 2019 (dashed line in Fig. 5) the BCM shows a reduction in $NO_2$ concentrations around Chinese New Year (5th February 2019), and it is likely that some reduction around the equivalent 2020 period (25th January 2020) would have occurred anyway. However, the 2020 reductions are significantly larger and more prolonged than in 2019. Similar to China, India shows large reductions in $NO_2$ concentration (60%) correlating with the implementation of restrictions in mid-March (Fig. 5); however, $NO_2$ concentrations have not yet recovered by the end of June, reflecting the prolonged duration of lockdown measures. Other areas of Asia, such as Hong Kong and Taipei, implemented smaller restrictions than China or India and they show significantly smaller decreases (less than 20%).

For Europe and the United States, we find widespread $NO_2$ reductions averaging 22% in March and 33% in April (Fig. 5). In some countries, recovery is evident as lockdown restrictions are removed or lessened (e.g., Greece, Romania) but in most countries, concentrations remain 20% or more below the business as usual scenario throughout May and June.

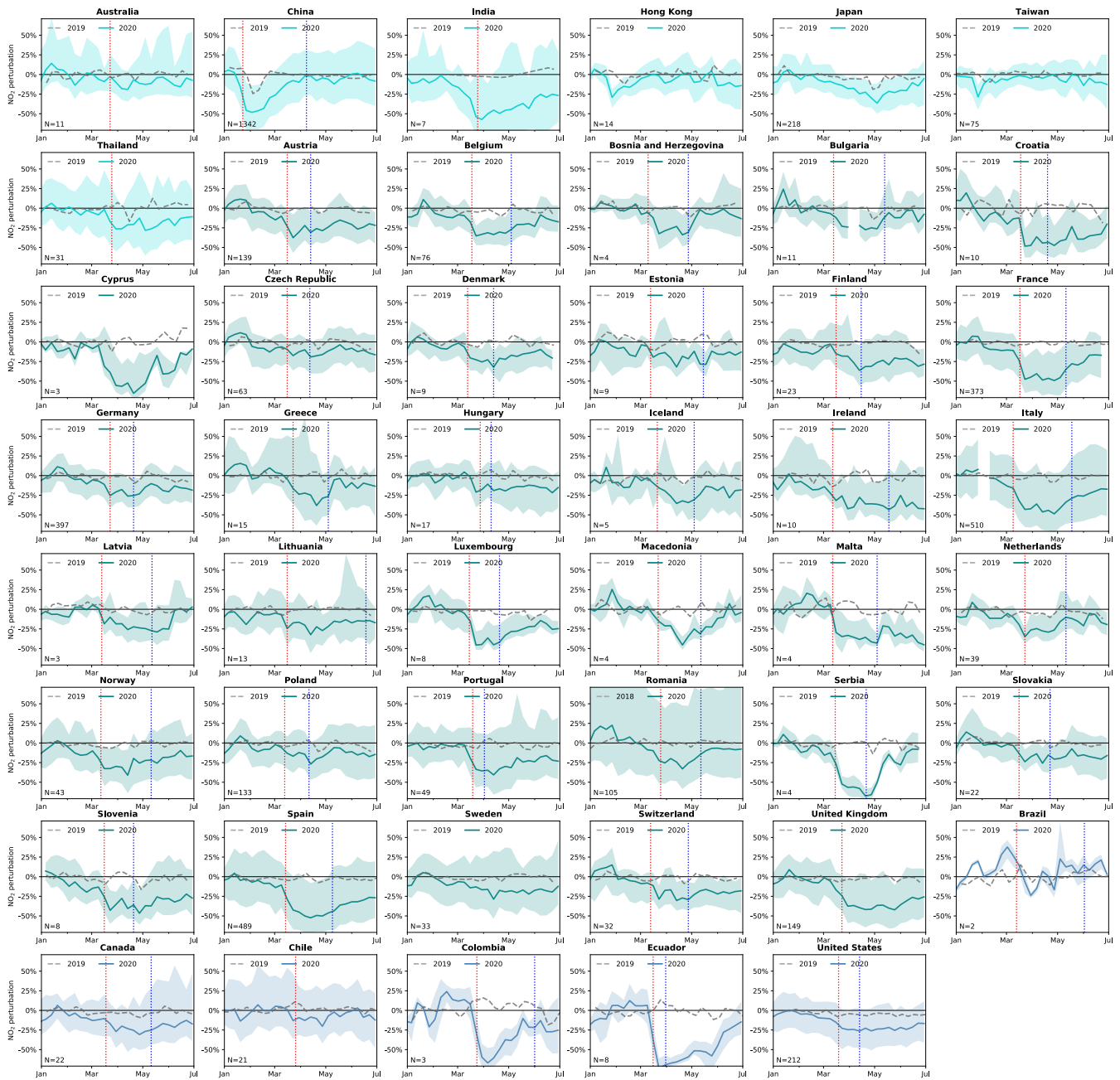

**Figure 5:** Seven-day average fractional difference between observed NO$_2$ and the BCM predictions for 46 countries between January 1 through June 30, aggregated from all sites across each country (number of sites in the bottom left of each panel). The thick line indicates the mean across all sites for the first half of 2020, with the shaded area representing the 5 to 95th percentile. Differing colours indicate differing regions (cyan: Asia & Australia; green: Europe; blue: Americas). The grey dashed line indicates the equivalent average for the same six month period in 2019 (2018 for countries with incomplete 2019 data). The red dashed vertical line indicates COVID-19 restriction dates, and the blue line indicates the beginning of easing measures.

**3.2 Ozone**

We follow the same methods for developing a business as usual counterfactual for $O_3$ as we did for $NO_2$ in section 3.1. Any change in $O_3$ concentration arising from COVID-19 restrictions is set against a large seasonal increase in concentrations in the Northern Hemisphere springtime (Fig. 6). This makes attributing changes in $O_3$ concentration more challenging than for $NO_2$. Our analysis shows an $O_3$ increase of up to 50% for some periods in cities with large $NO_2$ reductions (e.g., Wuhan, Milan, Quito; Fig. 3 and Fig. A4-A6), but there is much less convincing evidence for a systematic $O_3$ response across cities or on a regional level (Fig. 7). For example, our analysis shows little $O_3$ difference in Beijing and Madrid during lockdown despite $NO_2$ declines comparable to Wuhan or Milan (Fig. A4-A6). $O_3$ enhancements of 20% are found over Europe (e.g., Belgium, Italy, Luxembourg, Switzerland), with a peak in early April, approximately two weeks after lockdown started (Fig. 7).

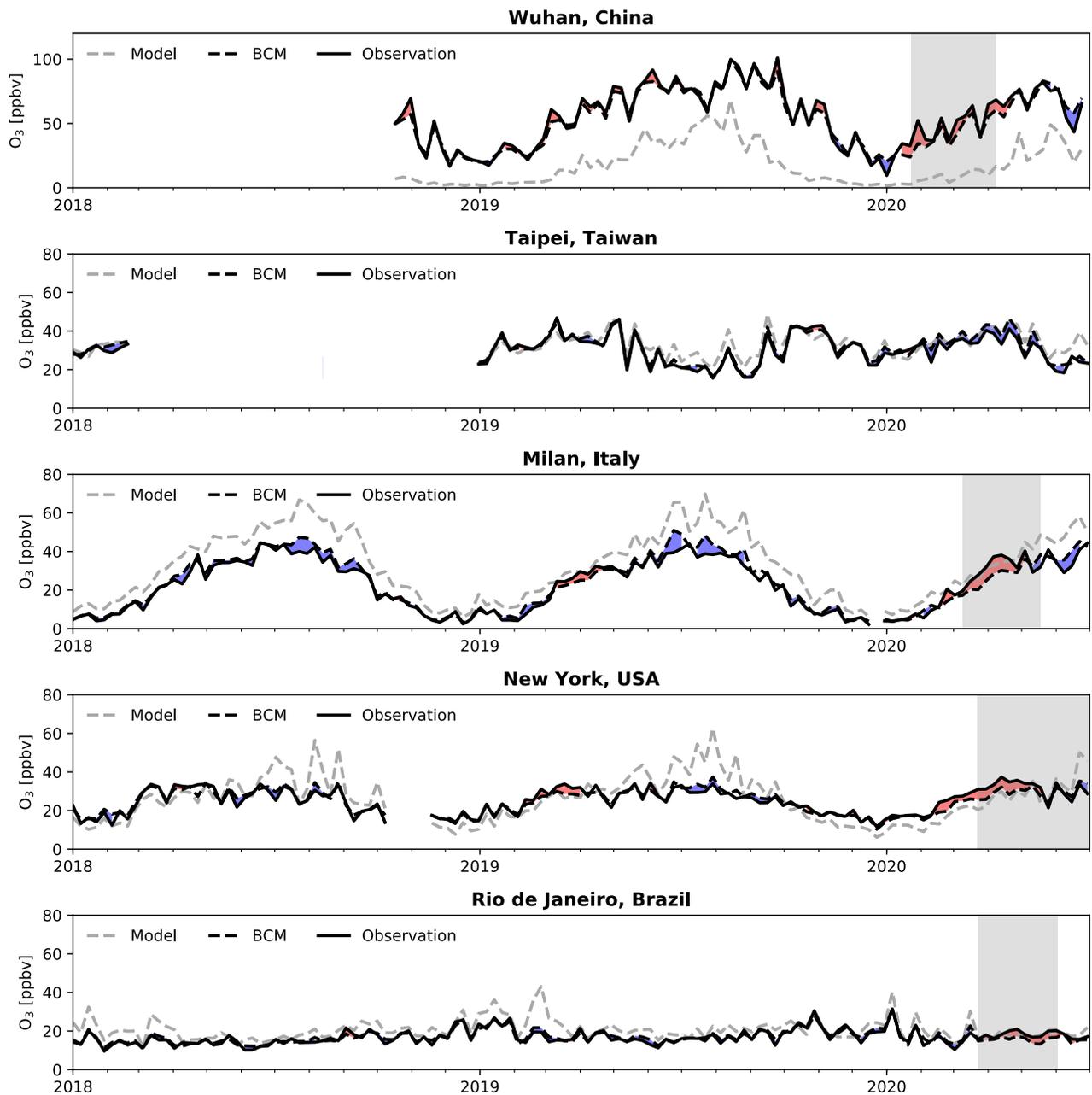

**Figure 6:** Comparison of O$_3$ surface concentrations for Wuhan, Taipei, Milan, New York, and Rio de Janeiro for January 2018 through June 2020. Observed values are shown in solid black, the original GEOS-CF model simulation is shown in dashed grey, and the BCM predictions are in dashed black. The area between observations and BCM predictions is shaded blue (red) if observations are lower (higher) than BCM predictions. The grey areas represent the period of lockdown. Shown are the 7-day average mean values for the 9, 18, 19, 14 and 4 observational sites in Wuhan, Taipei, Milan, New York, and Rio de Janeiro, respectively. Observations for China are only available starting in mid-September 2018.

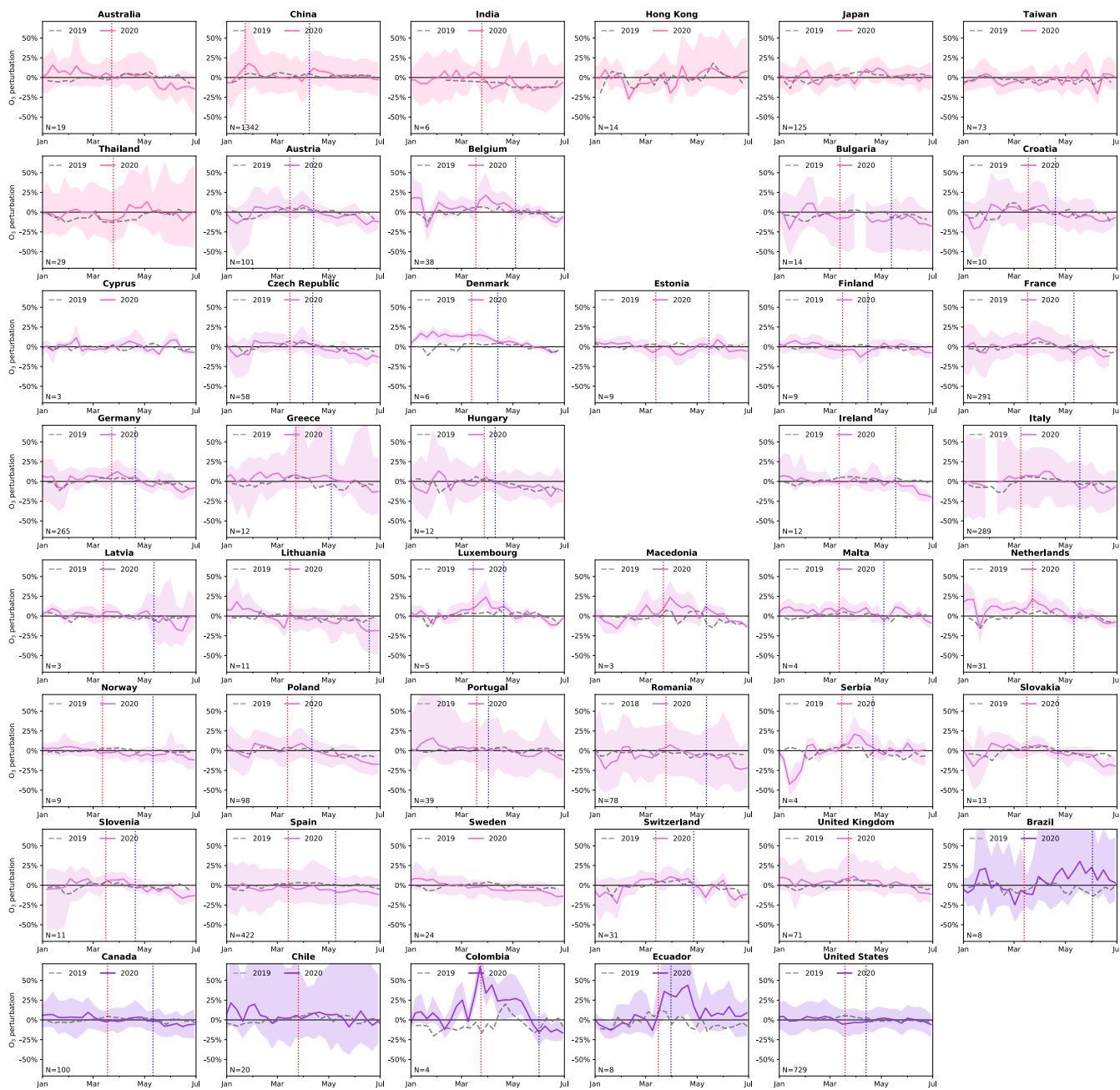

**Figure 7:** Similar to Figure 5 but without Bosnia and Herzegovina and Iceland. Differing colors indicate differing regions (pink: Asia & Australia; light purple: Europe; dark purple: Americas).

The analysis of $O_3$ is complicated by its nonlinear chemical response to $NO_x$ emissions. In the presence of sunlight, $O_3$ is produced chemically from the oxidation of volatile organic compounds in the presence of $NO_x$. Therefore, a decline in $NO_x$ emissions could decrease $O_3$ production and thus suppress $O_3$ concentrations. On the other hand, the process of $NO_x$ titration, in which freshly emitted NO rapidly reacts with $O_3$ to form $NO_2$, acts as a sink for $O_3$ (Seinfeld and Pandis, 2016). Odd oxygen ($O_x=NO_2+O_3$), is conserved when $O_3$ reacts with NO and thus offers a tool for separating these competing processes. Figure 8 presents the global mean diurnal cycle for $O_3$ and $O_x$ for the 5-month period since February 1, 2020 for both the observations and the BCM model. Compared to the BCM model, there has been an increase in the concentration of night time $O_3$ (midnight-5.00 local time, Fig. 8a) by 1 part per billion by volume (ppbv = nmol mol$^{-1}$) compared to the BCM, whereas $O_x$ shows a decrease of 1 ppbv (Fig. 8b). Thus, during the night, reduced NO emissions reduces $O_3$ titration, allowing $O_3$ concentrations to increase. During the afternoon, $O_3$ concentrations are lower by 1 ppbv (Fig. 8a), while observed $O_x$ concentrations are lower than the baseline model by almost 2 ppbv at 14:00 local time (Fig. 8b). We attribute the lower $O_x$ to reduced net $O_x$ production due to the lower $NO_x$ concentration, but as titration is also reduced, daytime $O_3$ concentrations are little changed. Overall changes to mean $O_3$ concentrations are small, but there is a flattening of the diurnal cycle.

As shown in the lower panels in Fig. 8, both factors - enhanced night time $O_3$ and reduced daytime $O_x$ - are more pronounced at locations where pre-existing $NO_2$ concentrations are high (> 15 ppbv). This suggests that the observed $O_3$ deviations from the BCM are indeed coupled to $NO_x$ reductions due to COVID-19 restrictions, given that those are most pronounced at polluted sites.

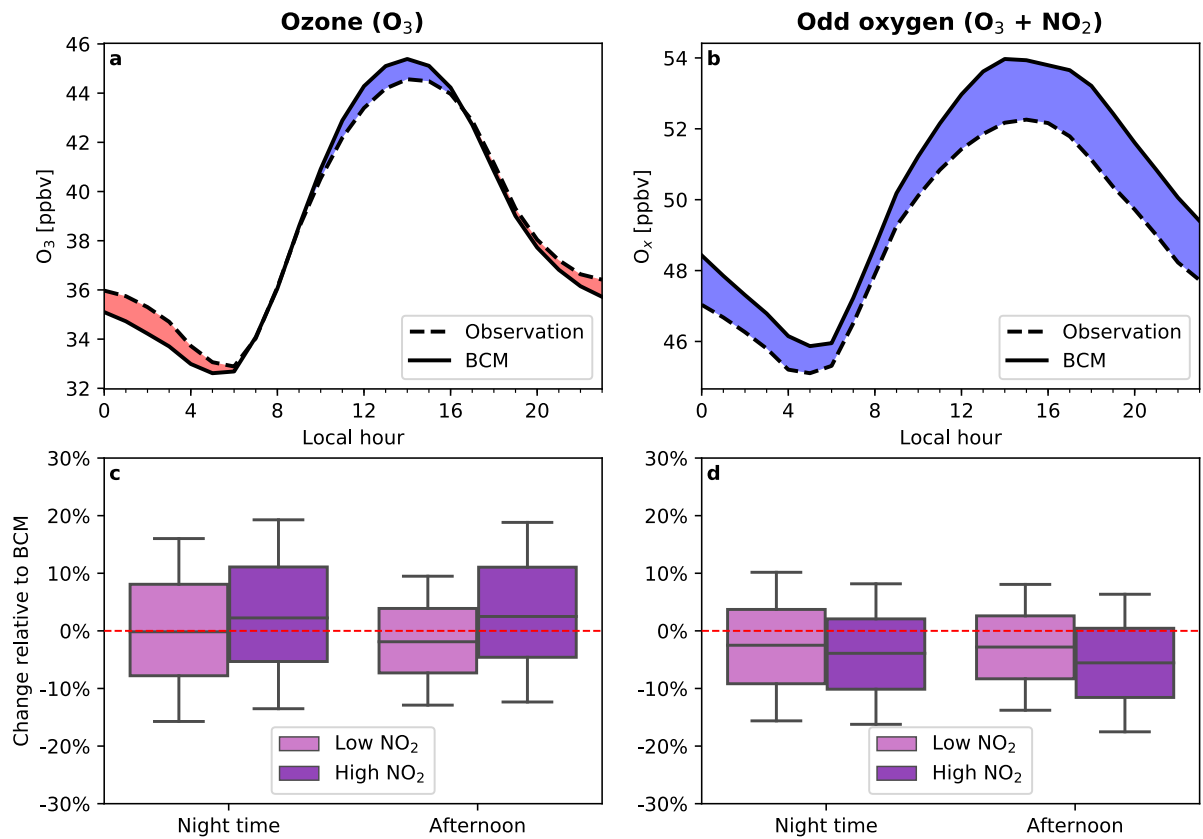

**Figure 8:** Observed and BCM modelled diurnal cycle of $O_3$ (a) and $O_x$ (b) from February 1, 2020 through June 30, 2020 with estimated corresponding changes in surface $O_3$ (c) and $O_x$ (d) relative to the BCM. Barplots (c and d) show observed changes during night time (0-5 local time) and the afternoon (12-17 local time) for locations with low (< 15 ppbv) and high (> 15 ppbv) $NO_2$ concentrations (based on the 2019 average).

### 3.3 $NO_x$ emission reductions

Changes in $NO_2$ concentrations do not necessarily reflect the same relative change in $NO_x$ emission (Lamsal et al., 2011). To analyze the response of $NO_2$ concentrations to a change in $NO_x$ emissions, we conducted a sensitivity simulation for the time period December 1, 2019 to June 8, 2020 using the GEOS-CF model with anthropogenic emissions scaled based on adjustment factors derived from $NO_2$ tropospheric columns observed by the NASA OMI instrument (Boersma et al., 2011). Daily scale factors were computed by normalizing coarse-resolution (2x2.5 degrees), 14-day $NO_2$ tropospheric column moving averages by the corresponding moving average for year 2018 (the emissions base year in GEOS-CF; section 2.2). Forest fire signals were filtered out based on QFED emissions and no scaling was applied over water. This results in anthropogenic emission adjustment factors of 0.3 to 1.4 (Fig. A7). These changes are of the same magnitude as obtained from the observation-BCM comparisons at cities globally (Fig. 5), thus this experiment provides the full range of experienced NO$_2$ changes. However, it should be noted that the scale factors do not necessarily coincide in space and time with the ones derived from observations and the BCM.

As shown in the left panel of Fig. 9, the GEOS-CF sensitivity simulation indicates that NO$_2$ concentrations drop, on average, by 80% of the fractional decrease in anthropogenic NO$_x$ emission, with a further diminishing effect for emission reductions greater than 50%. This reflects both the buffering effect of atmospheric chemistry and the presence of natural background concentrations. Based on bottom-up emissions estimates for 2015 from the Emission Database for Global Atmospheric Research (EDGAR v5.0_AP, Crippa et al., 2018, 2020) and using a constant concentration/emissions ratio of 0.8 based on the best fit line obtained from the model sensitivity simulation (dashed purple line in Fig. 9a), we calculate that the total reduction in anthropogenic NO$_x$ emissions due to COVID-19 containment measures during the first six months of 2020 amounted to 2.9 TgN (Fig. 9b and Table 2). This is equivalent to 5.1% of global annual anthropogenic NO$_x$ emissions (Table 2). Our estimate encompasses 46 countries that together account for 67% of the total emissions (excluding international shipping and aviation). We have no information for significant countries such as Russia, Indonesia, or anywhere in Africa due to the lack of publicly available near real-time air quality information. China accounts for the largest fraction of the reduced emissions (31%), followed by India (27%), the United States (20%), and Europe (13%).

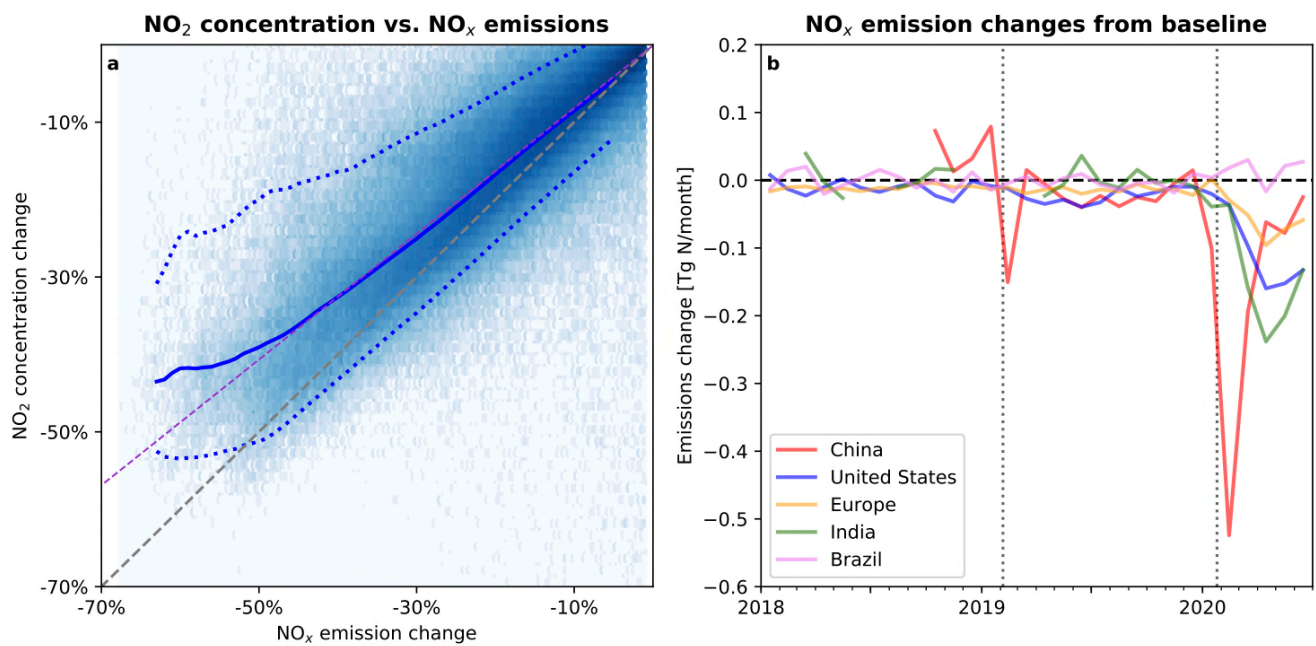

**Figure 9:** a) Response of NO$_2$ surface concentration (y-axis) to a change in NO$_x$ emissions (x-axis), as deduced from a model sensitivity simulation (see methods). The solid blue line shows the mean value across all individual grid cells (blue squares) and the dotted blue lines show the 5% and 95% quantiles. The dashed purple line shows the best linear fit. b) Estimated bi-weekly change in NO$_x$ emissions from the baseline since 2018 for China (red), United States (blue), Europe (yellow), India (green), and Brazil (purple). Dotted grey lines indicate Chinese New Year 2019 and 2020.

**Table 2:** Anthropogenic $NO_x$ emission reductions in GgN month$^{-1}$ as derived from $NO_2$ concentration changes.

| | Baseline[1] | Feb-20 | Mar-20 | Apr-20 | May-20 | Jun-20 |
|---|---|---|---|---|---|---|
| Australia | 52 | -2.8 | -2.5 | -9.0 | -5.1 | -6.6 |
| Austria | 6.1 | -0.4 | -1.7 | -2.0 | -1.5 | -1.7 |
| Belgium | 8.2 | -0.9 | -1.9 | -3.2 | -2.3 | -1.5 |
| Bosnia and Herzegovina | 2.7 | 0.0 | -0.5 | -0.9 | -0.3 | -0.2 |
| Brazil | 154 | 8 | 10 | -29 | 9 | n/a |
| Bulgaria | 3.8 | -0.1 | -0.6 | -1.2 | -0.7 | -0.4 |
| Canada | 63 | -5.4 | -10.8 | -19.1 | -18.2 | -13.4 |
| Chile | 17 | 0.2 | -0.2 | -2.5 | -1.3 | -0.8 |
| China | 990 | -524 | -194 | -62 | -78 | -25 |
| Colombia | 17 | 1.8 | -0.3 | -12.1 | -5.6 | -4.8 |
| Croatia | 2.0 | -0.2 | -0.6 | -1.0 | -0.9 | -0.8 |
| Cyprus | 0.7 | -0.1 | -0.2 | -0.5 | -0.3 | -0.2 |
| Czech Republic | 9.0 | -0.8 | -1.2 | -1.7 | -1.0 | -1.3 |
| Denmark | 4.0 | -0.3 | -0.6 | -1.3 | -0.8 | -0.7 |
| Ecuador | 11 | -1.4 | 1.1 | -3.7 | -8.9 | -7.4 |
| Estonia | 1.1 | -0.2 | -0.1 | -0.3 | -0.3 | -0.2 |
| Finland | 6.4 | -1.1 | -0.8 | -2.3 | -2.0 | -2.0 |
| France | 28 | -2.9 | -8.4 | -15.8 | -12.5 | -6.7 |
| Germany | 41 | -2.3 | -6.6 | -11.4 | -7.7 | -8.4 |
| Greece | 8.4 | 0.2 | -0.5 | -2.9 | -1.5 | -1.2 |
| Hong Kong | 7.5 | -1.6 | -0.2 | -0.4 | -0.3 | -1.2 |
| Hungary | 4.5 | -0.2 | -0.3 | -0.9 | -0.8 | -0.9 |
| Iceland | 0.2 | -0.03 | -0.02 | -0.07 | -0.06 | -0.05 |
| India | 391 | -36 | -159 | -238 | -200 | -133 |
| Ireland | 2.9 | -0.5 | -0.9 | -1.3 | -1.4 | -1.3 |
| Italy | 30 | -1.9 | -9.9 | -15.7 | -12.6 | -7.8 |
| Japan | 83 | -3.7 | -11.4 | -23.0 | -27.6 | -16.1 |
| Latvia | 1.2 | -0.03 | -0.08 | -0.33 | -0.38 | -0.10 |
| Lithuania | 1.7 | -0.17 | -0.27 | -0.50 | -0.38 | -0.32 |
| Luxembourg | 1.0 | 0.02 | -0.22 | -0.48 | -0.32 | -0.24 |
| Macedonia | 0.8 | 0.04 | -0.07 | -0.32 | -0.26 | -0.05 |
| Malta | 0.3 | 0.04 | -0.05 | -0.12 | -0.10 | -0.11 |
| Netherlands | 10 | -1.1 | -2.2 | -3.4 | -1.8 | -1.6 |
| Norway | 5.2 | -0.8 | -1.5 | -1.9 | -1.5 | -1.0 |
| Poland | 24 | -2.4 | -2.8 | -5.4 | -3.0 | -4.1 |
| Portugal | 5.8 | -0.3 | -1.1 | -2.5 | -1.8 | -1.4 |
| Romania | 8.5 | 0.6 | -1.2 | -2.7 | -1.5 | -0.8 |
| Serbia | 5.2 | -0.5 | -1.8 | -3.9 | -2.1 | -1.0 |
| Slovakia | 2.8 | -0.1 | -0.4 | -0.6 | -0.5 | -0.6 |
| Spain | 28 | -2.3 | -8.5 | -17.0 | -14.1 | -10.2 |
| Sweden | 7.1 | -0.6 | -0.9 | -1.7 | -1.7 | -1.5 |
| Switzerland | 3.0 | -0.1 | -0.5 | -0.9 | -0.8 | -0.8 |
| Taiwan | 31 | -3.7 | -1.4 | -1.1 | -1.7 | -3.8 |
| Thailand | 38 | -1.5 | -4.3 | -10.2 | -10.7 | -7.0 |
| United Kingdom | 33 | -2.0 | -5.3 | -15.9 | -15.7 | -12.1 |
| United States | 520 | -37 | -97 | -160 | -152 | -133 |
| Other countries[2] | 1340 | n/a | n/a | n/a | n/a | n/a |
| Shipping and Aviation | 671 | n/a | n/a | n/a | n/a | n/a |
| **Total** | **4681** | **-628** | **-532** | **-690** | **-594** | **-423** |

[1] EDGAR v5.0_AP 2015 annual emissions expressed as GgN month$^{-1}$ (Crippa et al., 2020)

[2] Primarily Indonesia, Iran, Mexico, Pakistan, Russia, Saudi Arabia, South Africa, South Korea, Vietnam

## 4 Conclusions

The combined interpretation of observations and model simulations using machine learning can be used to remove the compounding effect of meteorology and atmospheric chemistry, offering an effective tool to monitor and quantify changes in air pollution in near real-time. The global response to the COVID-19 pandemic presents a perfect testbed for this type of analysis, offering insights into the interconnectedness of human activity and air pollution. While national mitigation strategies have led to substantial regional $NO_2$ concentration decreases over the last decade in many places (e.g., Hilboll et al., 2013; Russell et al., 2013; Castellanos and Boersma, 2012), the widespread and near-instantaneous reduction in $NO_2$ following the implementation of COVID-19 containment measures indicates that there is still large potential to lower human exposure to $NO_2$ through reduction of anthropogenic $NO_x$ emissions.

The here derived $NO_2$ reductions are in good agreement with other emerging estimates. For instance, we determine an 18% decline over China for the 20 days after Chinese New Year relative to the preceding 20 days, consistent with the 21% reduction reported in Liu et al. (2020a). Similarly, our estimated 22% reduction over China for January to March 2020 is in excellent agreement with the 21-23% reported by Liu et al. (2020b). For Spain, we obtain an $NO_2$ reduction of 46% between March 14 to April 23, again in close agreement with the values reported in Petetin et al. (2020).

The response of $O_3$ to $NO_2$ declines in the wake of the COVID-19 outbreak is complicated by the competing influences of atmospheric chemistry. From February through June 2020, the diurnal observation-BCM comparisons suggest that the reduction in photochemical production was offset by a smaller loss from titration. This resulted in a flattening of the diurnal cycle and an insignificant net change in surface $O_3$ over a diurnal cycle. The competing impacts of reduced $NO_x$ emissions on $O_3$ production and loss are dependent on the local chemical and meteorological environment. This is reflected in the variable geographical response of $O_3$ following the implementation of COVID-19 restrictions (Le et al., 2020; Dantas et al., 2020). Moreover, as atmospheric reactivity increases through the Northern Hemisphere spring and summer, the relative importance of photochemical production is expected to increase in the Northern Hemisphere.

To assess the potential seasonal-scale impact of reduced anthropogenic emissions on $O_3$, we conducted two free-running forecast simulations between June 8 through August 31, 2020, initialized from the GEOS-CF simulation and the sensitivity simulation described in Section 3.3, respectively. Both simulations use the same biomass burning emissions based on a historical QFED climatology. For the forecast sensitivity experiment, we assume a sustained reduction in global anthropogenic emissions of $NO_x$, carbon monoxide (CO), and VOCs. As shown in Fig. 10, this leads to a general decrease in $O_3$ concentrations of 10-20% over Eastern China, Europe, and the Western and Northeast US in July and August 2020 relative to the business-as-usual reference forecast simulation. However, it is also notable that in some locations the model forecast $O_3$ concentrations increase by an equivalent amount (e.g., Scandinavia, South Central US and Mexico, Northern

India), reflecting the high nonlinearity of atmospheric chemistry. This highlights the complex interactions between emissions, chemistry, and meteorology and their impact on air pollution on different time scales. While large reductions in $NO_2$ concentrations are achievable and immediately follow curtailments in $NO_x$ emissions, the $O_3$ response is more complicated and can be in the opposite direction, at times by as much as 50% (Jhun et al., 2015, Le et al., 2020). This shows the difficulties faced by policy makers in curbing $O_3$ pollution.

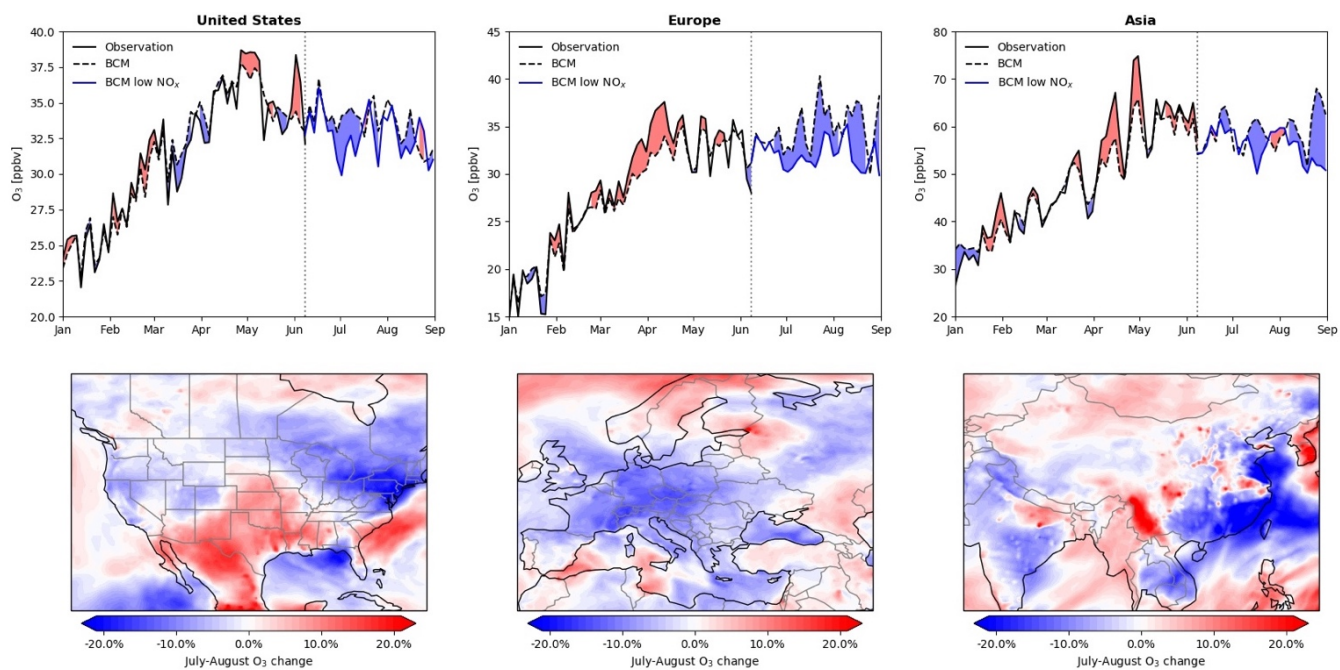

**Figure 10:** Change in mean surface $O_3$ over the United States, Europe, and Asia for a sensitivity simulation with altered anthropogenic emissions. Top panels show daily average $O_3$ concentrations at all observation sites within the given region (solid black, Jan-Jun), the bias-corrected GEOS-CF model ("BCM", solid black, Jan 1st - Jun 8th) continued with a business as usual GEOS-CF forecast from Jun 9th - Aug 31st, and GEOS-CF forecast assuming sustained 20% anthropogenic emission reduction (blue). Bottom panels show mean changes in surface $O_3$ for July and August for the low emissions simulation relative to the business as usual forecast.

## Appendix

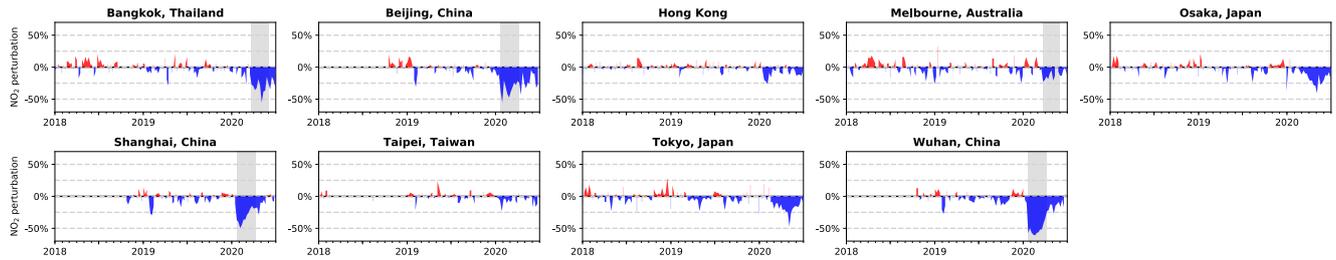

**Figure A1:** Normalized fractional NO$_2$ perturbations (observation - bias-corrected model, normalized by the bias-corrected model prediction) from Jan 1, 2018 through June 2020 for 9 cities in Asia and Australia. Grey shaded areas indicate COVID-19 lockdown periods.

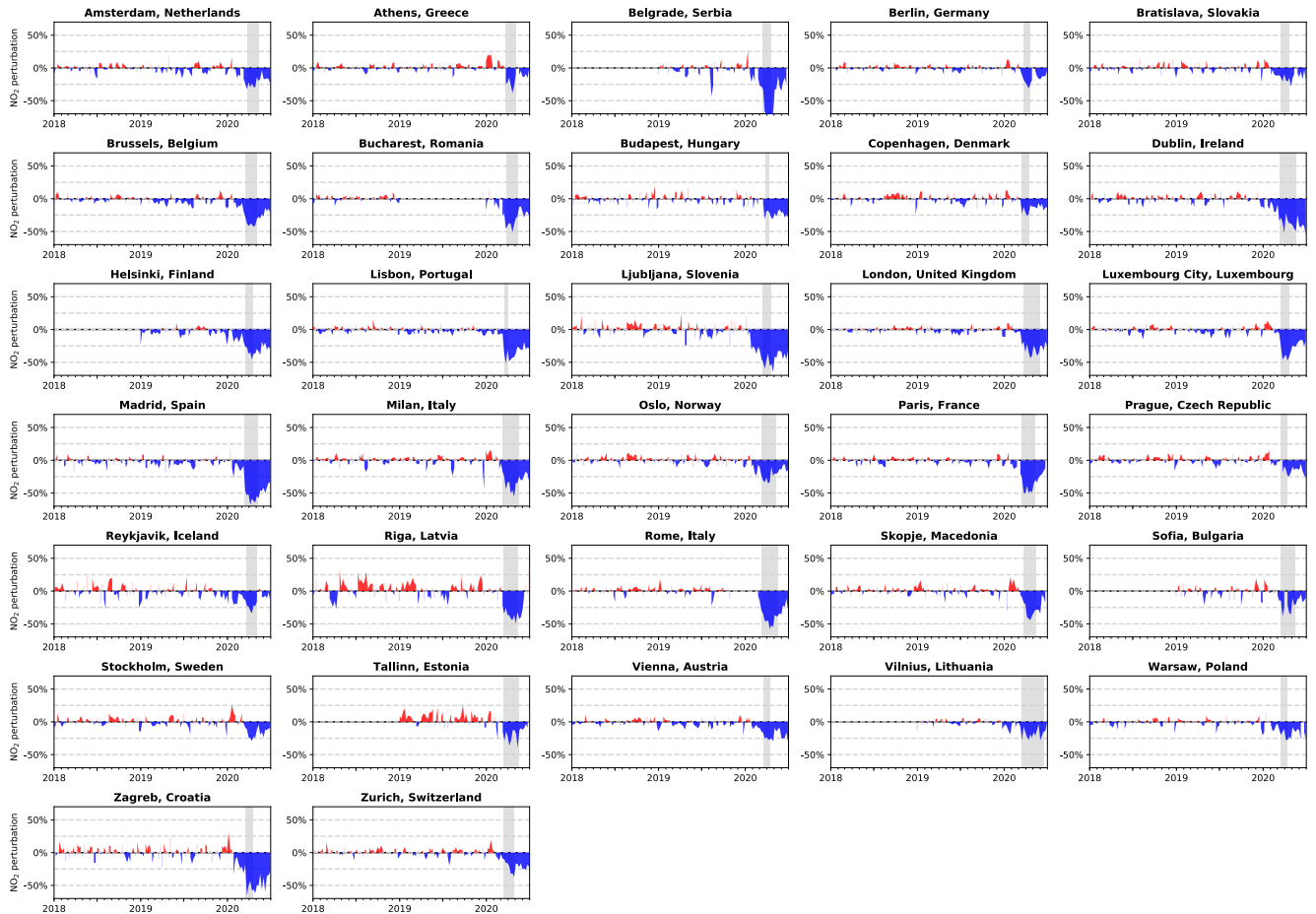

**Figure A2:** Normalized fractional NO$_2$ perturbations (observation - bias-corrected model, normalized by the bias-corrected model prediction) from Jan 1, 2018 through June 2020 for 32 cities in Europe. Grey shaded areas indicate COVID-19 lockdown periods.

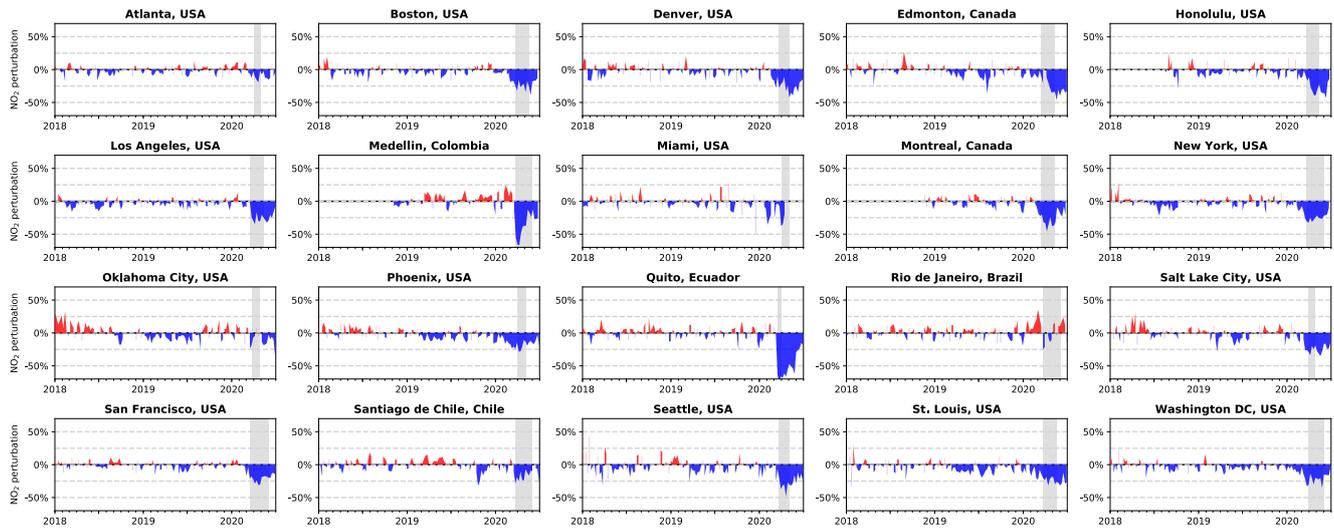

**Figure A3:** Normalized fractional NO$_2$ perturbations (observation - bias-corrected model, normalized by the bias-corrected model prediction) from Jan 1, 2018 through June 2020 for 20 cities in North and South America. Grey shaded areas indicate COVID-19 lockdown periods.

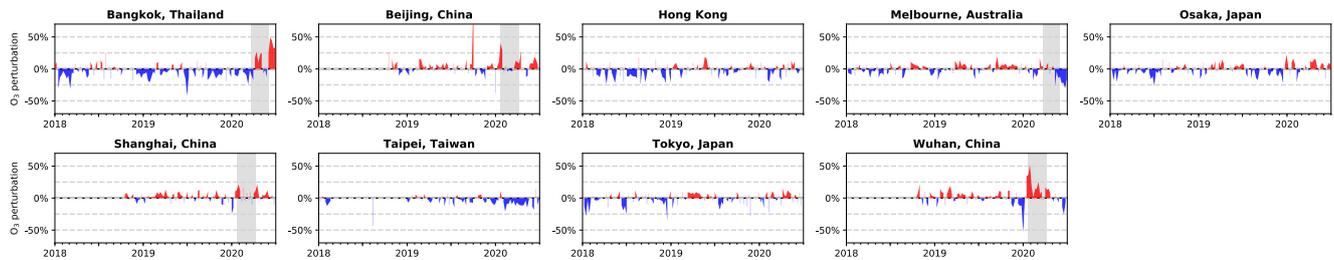

**Figure A4:** Normalized fractional O$_3$ perturbations (observation - bias-corrected model, normalized by the bias-corrected model prediction) from Jan 1, 2018 through June 2020 for 9 cities in Asia and Australia. Grey shaded areas indicate COVID-19 lockdown periods.

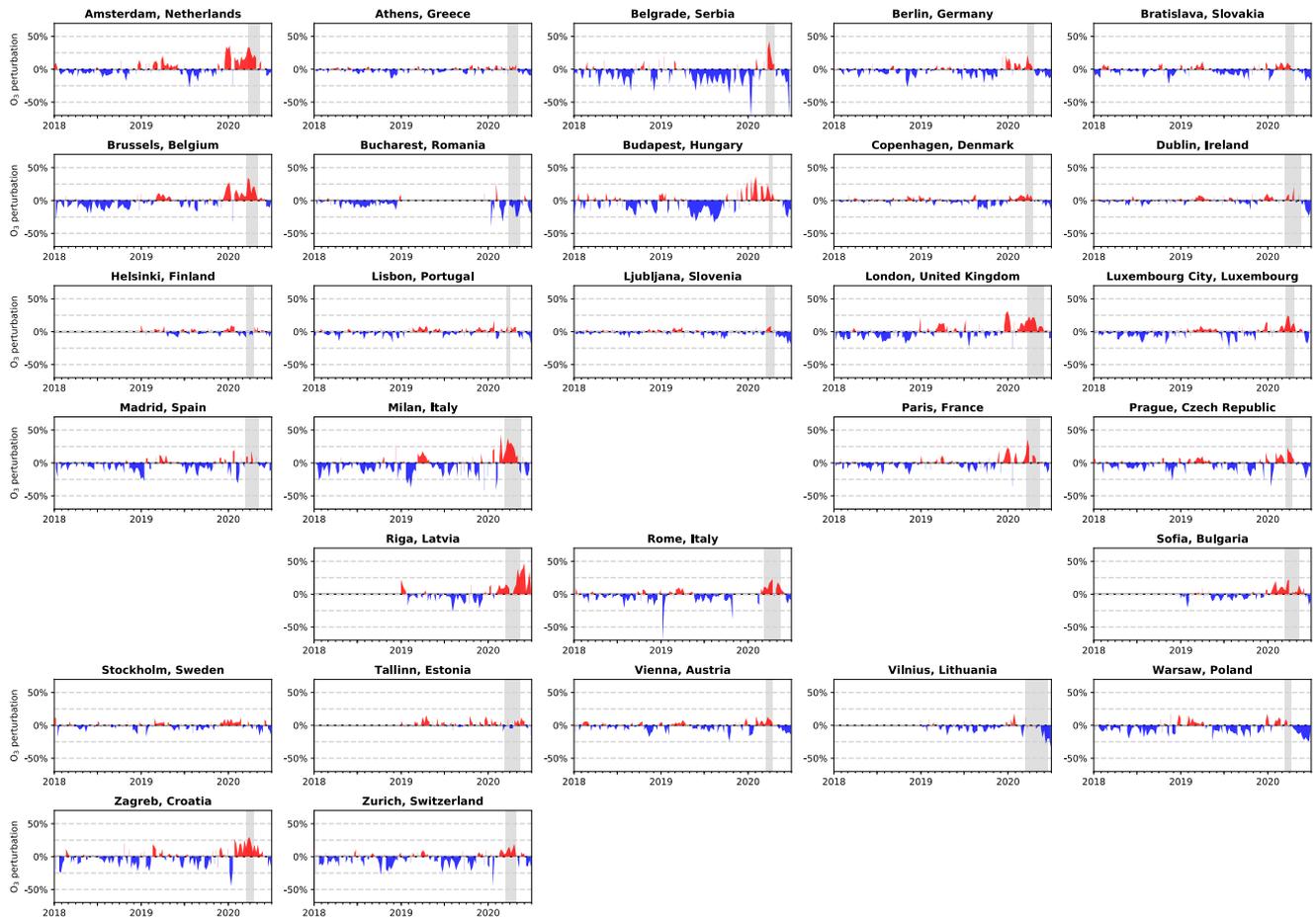

**Figure A5:** Normalized fractional O$_3$ perturbations (observation - bias-corrected model, normalized by the bias-corrected model prediction) from Jan 1, 2018 through June 2020 for 29 cities in Europe. Grey shaded areas indicate COVID-19 lockdown periods. No observations are available for Reykjavik, Oslo, and Skopje.

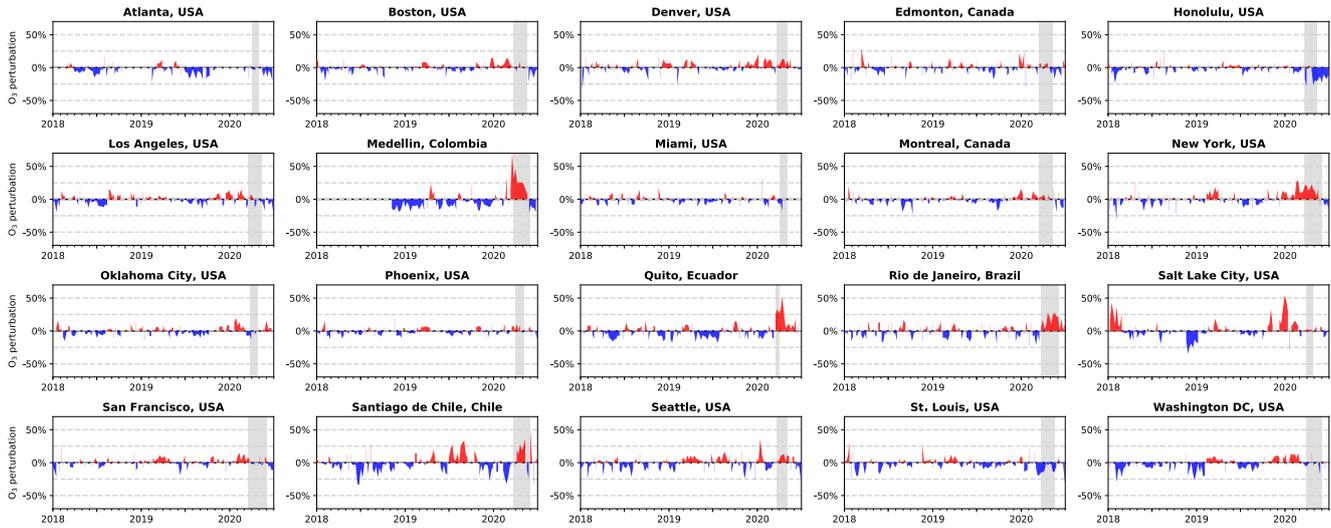

**Figure A6:** Normalized fractional $O_3$ perturbations (observation - bias-corrected model, normalized by the bias-corrected model prediction) from Jan 1, 2018 through June 2020 for 20 cities in North and South America. Grey shaded areas indicate COVID-19 lockdown periods.

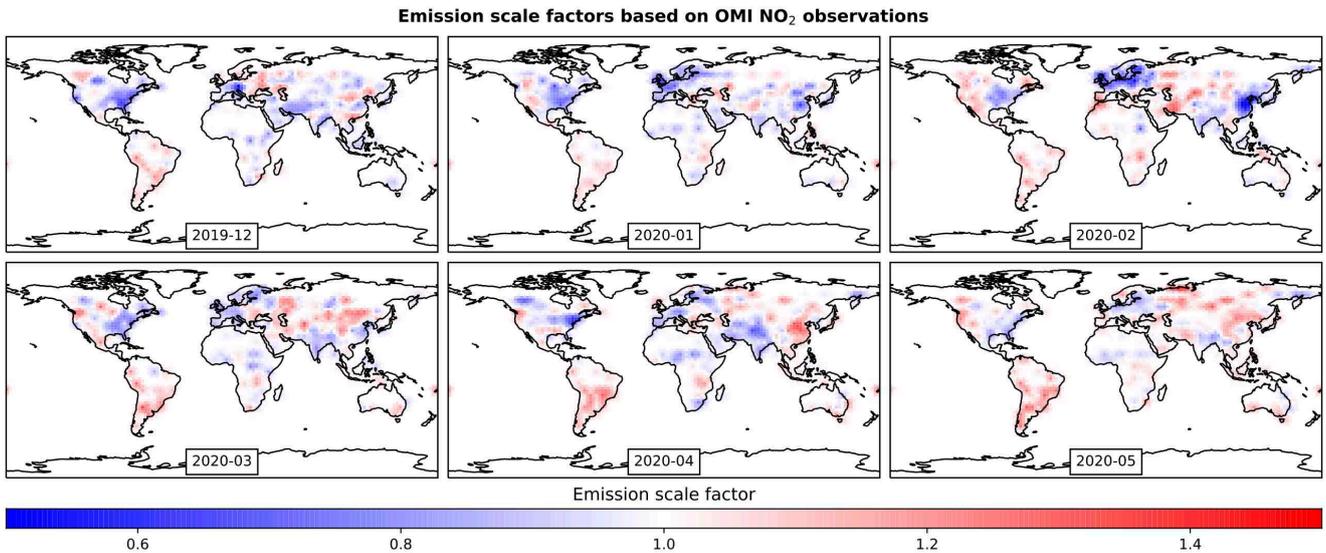

**Figure A7:** Emission scale factors used for model sensitivity simulation. Shown are the monthly average perturbations applied to the 2018 anthropogenic base emissions.

**Table A1:** Monthly changes in $NO_2$ concentrations relative to the bias-corrected model predictions for cities in Asia & Australia.

|  | Jan-20 | Feb-20 | Mar-20 | Apr-20 | May-20 | Jun-20 |
|---|---|---|---|---|---|---|
| Bangkok, Thailand | -4.9% | -8.2% | -16.5% | -26.1% | -38.1% | -24.5% |
| Beijing, China | -20.9% | -38.1% | -31.0% | -23.4% | -23.6% | -18.6% |
| Hong Kong | -6.7% | -14.1% | -3.4% | -3.9% | -4.1% | -13.3% |
| Melbourne, Australia | 5.4% | 0.0% | -7.1% | -14.1% | -11.1% | -8.1% |
| Osaka, Japan | -10.3% | -2.5% | -10.0% | -23.1% | -25.6% | -14.1% |
| Shanghai, China | -13.8% | -40.6% | -24.3% | -20.8% | -6.3% | -2.4% |
| Taipei, Taiwan | -7.2% | -6.0% | -5.0% | -0.4% | -2.6% | -10.7% |
| Tokyo, Japan | 1.1% | -2.7% | -13.9% | -21.4% | -26.8% | -11.7% |
| Wuhan, China | -19.5% | -56.8% | -51.8% | -22.5% | -14.5% | -10.0% |

**Table A2:** Monthly changes in $NO_2$ concentrations relative to the bias-corrected model predictions for cities in Europe.

|  | 20-Jan | 20-Feb | 20-Mar | 20-Apr | 20-May | 20-Jun |
|---|---|---|---|---|---|---|
| Amsterdam, Netherlands | 2.0% | -9.3% | -15.3% | -26.8% | -16.4% | -17.0% |
| Athens, Greece | 15.7% | 3.5% | -4.1% | -25.7% | -7.1% | -11.9% |
| Belgrade, Serbia | 1.2% | -10.5% | -39.7% | -72.5% | -30.8% | -24.0% |
| Berlin, Germany | 7.6% | -3.6% | -6.9% | -23.9% | -9.7% | -13.3% |
| Bratislava, Slovakia | 7.8% | -9.6% | -14.8% | -16.1% | -9.7% | -10.3% |
| Brussels, Belgium | -3.8% | -10.0% | -24.4% | -40.3% | -29.4% | -18.6% |
| Bucharest, Romania | -1.3% | -11.0% | -22.9% | -40.6% | -22.9% | -22.4% |
| Budapest, Hungary | -0.1% | -6.4% | -10.5% | -21.6% | -22.7% | -23.7% |
| Copenhagen, Denmark | 7.1% | -2.2% | -9.5% | -17.9% | -11.3% | -13.0% |
| Dublin, Ireland | -10.3% | -11.6% | -28.0% | -36.1% | -37.9% | -42.4% |
| Helsinki, Finland | -9.9% | -15.5% | -18.9% | -38.3% | -31.9% | -26.1% |
| Lisbon, Portugal | -5.3% | -3.2% | -27.1% | -41.2% | -32.6% | -27.6% |
| Ljubljana, Slovenia | -4.8% | -30.4% | -46.6% | -48.2% | -40.0% | -39.8% |
| London, United Kingdom | 0.0% | -0.4% | -14.2% | -34.2% | -32.7% | -25.2% |
| Luxembourg City, Luxembourg | 6.8% | -1.0% | -23.5% | -39.7% | -25.8% | -18.7% |
| Madrid, Spain | -0.6% | -10.2% | -31.5% | -60.9% | -52.5% | -40.8% |
| Milan, Italy | 11.1% | -1.0% | -23.8% | -41.3% | -33.8% | -23.4% |
| Oslo, Norway | -1.6% | -13.5% | -23.0% | -25.2% | -19.5% | -11.6% |
| Paris, France | -0.3% | -10.1% | -25.8% | -45.7% | -33.7% | -18.3% |
| Prague, Czech Republic | 9.2% | -6.0% | -8.0% | -15.5% | -13.9% | -15.3% |
| Reykjavik, Iceland | -12.3% | -11.8% | -13.8% | -26.8% | -6.9% | -7.7% |
| Riga, Latvia | 1.6% | -1.8% | -18.0% | -40.3% | -40.5% | -4.0% |
| Rome, Italy | N/A | -0.7% | -32.7% | -50.2% | -38.7% | -23.2% |
| Skopje, Macedonia | -3.6% | 10.1% | -7.1% | -36.7% | -31.4% | -15.2% |
| Sofia, Bulgaria | 9.4% | -0.6% | -17.0% | -28.1% | -20.4% | -11.8% |
| Stockholm, Sweden | 8.4% | 2.6% | -4.5% | -23.8% | -15.4% | -14.0% |
| Tallinn, Estonia | 6.6% | -6.0% | -11.0% | -22.8% | -19.6% | -6.6% |
| Vienna, Austria | 2.7% | -9.3% | -16.9% | -25.0% | -13.2% | -23.3% |
| Vilnius, Lithuania | -9.7% | -3.5% | -10.9% | -22.7% | -14.5% | -17.7% |
| Warsaw, Poland | -2.9% | -9.3% | -12.5% | -22.5% | -14.3% | -14.1% |
| Zagreb, Croatia | 5.3% | -17.2% | -34.8% | -51.9% | -48.6% | -40.9% |
| Zurich, Switzerland | 8.9% | -5.6% | -10.5% | -26.3% | -20.0% | -23.3% |

**Table A3:** Monthly changes in $NO_2$ concentrations relative to the bias-corrected model predictions for cities in North and South America.

|  | Jan-20 | Feb-20 | Mar-20 | Apr-20 | May-20 | Jun-20 |
|---|---|---|---|---|---|---|
| Atlanta, USA | 8.6% | 0.5% | -2.5% | -12.1% | -7.5% | -6.7% |
| Boston, USA | -6.2% | -2.3% | -20.1% | -24.4% | -27.0% | -14.6% |
| Denver, USA | -2.7% | -6.3% | -19.4% | -21.2% | -36.0% | -23.3% |
| Edmonton, Canada | -2.6% | -10.4% | -12.4% | -17.9% | -34.8% | -33.4% |
| Honolulu, USA | -1.8% | -1.7% | -10.4% | -28.0% | -30.5% | -28.5% |
| Los Angeles, USA | 5.1% | -3.7% | -9.8% | -28.0% | -27.5% | -22.1% |
| Medellin, Colombia | -2.6% | 13.2% | -9.1% | -53.4% | -24.7% | -23.7% |
| Miami, USA | -3.6% | -24.2% | -10.7% | -33.2% | N/A | N/A |
| Montreal, Canada | -2.0% | -0.5% | -20.7% | -34.3% | -26.0% | -16.0% |
| New York, USA | -5.5% | -8.2% | -20.5% | -28.4% | -25.2% | -16.9% |
| Oklahoma City, USA | 2.2% | -1.9% | -5.7% | -2.4% | -14.0% | -8.8% |
| Phoenix, USA | -8.2% | -12.6% | -20.2% | -21.7% | -13.9% | -12.1% |
| Quito, Ecuador | -16.9% | -2.8% | -40.4% | -67.3% | -57.8% | -23.3% |
| Rio de Janeiro, Brazil | -2.8% | 3.9% | -0.5% | -16.3% | 4.1% | 14.0% |
| Salt Lake City, USA | -2.4% | -3.9% | -16.6% | -25.0% | -28.3% | -20.0% |
| San Francisco, USA | 5.5% | -1.5% | -15.2% | -27.3% | -19.6% | -16.6% |
| Santiago de Chile, Chile | -2.9% | -5.4% | -6.4% | -20.9% | -12.1% | -6.4% |
| Seattle, USA | -9.1% | -3.0% | -13.5% | -34.6% | -27.1% | -19.7% |
| St. Louis, USA | -4.9% | -12.3% | -16.2% | -24.8% | -24.4% | -24.9% |
| Washington DC, USA | -5.9% | -10.5% | -18.9% | -21.8% | -25.6% | -14.8% |

**Table A4:** Monthly changes in $O_3$ concentrations relative to the bias-corrected model predictions for cities in Asia & Australia.

|  | Jan-20 | Feb-20 | Mar-20 | Apr-20 | May-20 | Jun-20 |
|---|---|---|---|---|---|---|
| Bangkok, Thailand | -8.1% | -3.2% | -10.7% | 11.6% | 3.7% | 38.1% |
| Beijing, China | 9.9% | -0.1% | 1.8% | 12.6% | -3.6% | 11.4% |
| Hong Kong | 0.8% | -8.5% | -2.0% | -2.6% | 9.4% | -3.1% |
| Melbourne, Australia | -0.8% | -8.7% | 3.4% | 1.1% | -6.9% | -21.0% |
| Osaka, Japan | 12.4% | -4.0% | -0.8% | 11.2% | 3.5% | 2.5% |
| Shanghai, China | -1.3% | 7.3% | 0.8% | 10.6% | 2.7% | 1.7% |
| Taipei, Taiwan | -0.7% | -9.9% | -7.9% | -9.4% | -6.4% | -6.5% |
| Tokyo, Japan | -4.3% | 2.4% | -1.4% | 6.4% | 4.2% | -0.3% |
| Wuhan, China | 13.6% | 16.9% | 9.5% | 11.8% | -2.0% | -8.8% |

Table A5: Monthly changes in $O_3$ concentrations relative to the bias-corrected model predictions for cities in Europe.

|  | Jan-20 | Feb-20 | Mar-20 | Apr-20 | May-20 | Jun-20 |
|---|---|---|---|---|---|---|
| Amsterdam, Netherlands | 18.9% | 13.3% | 19.5% | 24.1% | 4.4% | -5.2% |
| Athens, Greece | 0.7% | 3.7% | -0.6% | 3.4% | -3.7% | -5.5% |
| Belgrade, Serbia | -25.4% | -0.3% | 12.4% | 13.4% | -7.0% | -16.3% |
| Berlin, Germany | 6.5% | 7.3% | 6.8% | 1.7% | -5.5% | -9.4% |
| Bratislava, Slovakia | -7.6% | 7.4% | 5.7% | 0.5% | -2.5% | -11.2% |
| Brussels, Belgium | 9.1% | 10.2% | 15.6% | 18.7% | 2.6% | -6.3% |
| Bucharest, Romania | -12.9% | -0.1% | -11.9% | -9.8% | -13.7% | -4.9% |
| Budapest, Hungary | 16.7% | 17.8% | 14.7% | 4.7% | -2.1% | -15.9% |
| Copenhagen, Denmark | 2.9% | 4.0% | 6.8% | 2.1% | -1.4% | -5.3% |
| Dublin, Ireland | 4.9% | -2.2% | 2.4% | 8.0% | -1.7% | -13.9% |
| Helsinki, Finland | 7.1% | -2.9% | -3.9% | -1.8% | 3.1% | -3.3% |
| Lisbon, Portugal | 6.5% | -4.1% | 2.2% | 4.8% | -6.1% | -3.2% |
| Ljubljana, Slovenia | -3.6% | 0.0% | -0.4% | 2.4% | -6.7% | -12.3% |
| London, United Kingdom | 10.9% | 4.8% | 14.0% | 20.0% | 6.7% | -2.5% |
| Luxembourg City, Luxembourg | 2.7% | 3.2% | 12.8% | 11.2% | -0.1% | -9.6% |
| Madrid, Spain | 0.9% | -10.0% | 2.9% | 6.0% | -2.6% | -6.1% |
| Milan, Italy | 4.5% | 11.0% | 19.6% | 24.6% | -0.9% | -11.3% |
| Paris, France | 5.3% | 3.4% | 14.0% | 8.4% | 1.6% | -8.2% |
| Prague, Czech Republic | -11.8% | 10.7% | 9.8% | 7.8% | -1.3% | -12.7% |
| Riga, Latvia | 2.7% | 6.5% | 12.9% | 2.7% | 34.1% | 24.2% |
| Rome, Italy | N/A | -6.2% | 7.9% | 14.2% | 10.8% | -6.7% |
| Sofia, Bulgaria | 5.3% | 10.9% | 10.9% | 3.4% | 4.9% | -4.1% |
| Stockholm, Sweden | 6.0% | 6.2% | 0.8% | -2.8% | -7.9% | -1.0% |
| Tallinn, Estonia | 3.8% | 2.3% | 0.3% | -1.7% | 6.9% | -0.3% |
| Vienna, Austria | -7.9% | 7.5% | 5.1% | 4.4% | -4.1% | -10.5% |
| Vilnius, Lithuania | 3.7% | -0.1% | -2.4% | -0.5% | -5.6% | -20.3% |
| Warsaw, Poland | 1.7% | 8.5% | 1.9% | -2.2% | -12.7% | -20.9% |
| Zagreb, Croatia | -13.2% | 17.2% | 18.5% | 16.6% | 7.6% | -5.9% |
| Zurich, Switzerland | -12.2% | 3.0% | 6.6% | 12.9% | -2.6% | -7.7% |

Table A6: Monthly changes in $O_3$ concentrations relative to the bias-corrected model predictions for cities in North and South America.

|  | Jan-20 | Feb-20 | Mar-20 | Apr-20 | May-20 | Jun-20 |
|---|---|---|---|---|---|---|
| Atlanta, USA | N/A | 4.5% | -3.8% | -2.4% | -4.4% | -8.0% |
| Boston, USA | 4.9% | 7.3% | 4.8% | 0.7% | -4.7% | -7.1% |
| Denver, USA | 7.5% | 12.7% | 0.6% | 8.1% | 6.3% | -0.5% |
| Edmonton, Canada | 2.9% | 3.4% | -1.9% | 4.2% | -7.0% | -2.9% |
| Honolulu, USA | 0.5% | -5.9% | -8.7% | -7.2% | -18.9% | -15.8% |
| Los Angeles, USA | 1.7% | 10.1% | -2.6% | -3.5% | -6.6% | -4.3% |
| Medellin, Colombia | 2.2% | -1.5% | 25.6% | 29.5% | 10.7% | -12.8% |
| Miami, USA | 8.8% | -5.3% | 0.9% | -9.0% | N/A | N/A |
| Montreal, Canada | 5.0% | 5.3% | 3.0% | 4.4% | -2.1% | -8.0% |
| New York, USA | 5.2% | 16.9% | 14.6% | 16.3% | 7.4% | -3.6% |
| Oklahoma City, USA | 6.7% | 9.0% | -3.4% | -3.2% | 1.1% | 5.2% |
| Phoenix, USA | -3.3% | 2.8% | 2.0% | 3.8% | 1.1% | -1.9% |
| Quito, Ecuador | -7.4% | -0.2% | 11.1% | 39.2% | 9.7% | 7.7% |
| Rio de Janeiro, Brazil | 13.3% | -6.7% | 0.9% | 24.0% | 26.2% | 10.3% |
| Salt Lake City, USA | 24.9% | 10.0% | -2.3% | 1.1% | 2.9% | -3.2% |
| San Francisco, USA | -1.4% | 9.7% | 6.7% | -2.5% | 2.3% | -4.3% |
| Santiago de Chile, Chile | -4.8% | -15.7% | -7.8% | 14.5% | 8.9% | 6.0% |
| Seattle, USA | 11.4% | -10.1% | 1.2% | 7.7% | 4.8% | -4.7% |
| St. Louis, USA | -6.1% | 8.6% | -14.2% | -7.2% | -7.0% | -0.2% |
| Washington DC, USA | 8.3% | 11.1% | -1.5% | 0.9% | -3.8% | -1.4% |

*Data availability.* The model output and air quality observations used in this study are all publicly available (see methods). The output from the GEOS-CF sensitivity simulation as well as the bias-corrected model predictions are available from CAK per request.

*Author contributions.* CAK and MJE designed the study and conducted the main analyses. CAH and SM contributed OpenAQ observations. TO provided observations and interpretations for Japan. FCM and BBF provided observations and interpretations for Rio de Janeiro, and MVDS provided observations and interpretations for Quito. RGR provided observations for Melbourne and helped analyze results for Australia. KEK and RAL conducted the GEOS-CF simulations. KEK and CAK conducted the GEOS-CF sensitivity experiments and forecasts. LHF conducted $NO_x$ to $NO_2$ sensitivity simulations. SP contributed to overall study design and context discussion. All authors contributed to the writing.

*Competing interests.* The authors declare that they have no conflict of interest.

*Acknowledgements.* Resources supporting the model simulations were provided by the NASA Center for Climate Simulation at the Goddard Space Flight Center (https://www.nccs.nasa.gov/services/discover). We thank Jenny Fisher (U. Wollongong, Australia) for helpful discussions. CAK, KEK and SP acknowledge support by the NASA Modeling, Analysis and Prediction (MAP) Program. MJE and LHF are thankful for support from the University of York's Viking, HPC facility.